\documentclass[aps,twocolumn,preprintnumbers,amsmath,amssymb,floatfix]{revtex4}

\usepackage{graphicx}
\usepackage{epsfig}

\newcommand{\comment}[1]{}
\newcommand{\thold}{T_{\mbox{\small hold}}}
\newcommand{\tinstab}{\tau_{FM}}
\newcommand{\kinstab}{k_{FM}}
\newcommand{\rhovec}{\vec{\rho}}

\begin{document}
\title{Spontaneous symmetry breaking in a quenched ferromagnetic spinor Bose condensate}

\author{L. E. Sadler, J. M. Higbie, S. R. Leslie, M. Vengalattore and D. M. Stamper-Kurn}

\affiliation{Department of Physics, University of California,
Berkeley CA 94720}

\maketitle

\textbf{A central goal in condensed matter and modern atomic physics
is the exploration of many-body quantum phases and the universal
characteristics of quantum phase transitions in so far as they
differ from those established for thermal phase transitions.
Compared with condensed-matter systems, atomic gases are more
precisely constructed and also provide the unique opportunity to
explore quantum dynamics far from equilibrium. Here we identify a
second-order quantum phase transition in a gaseous spinor
Bose-Einstein condensate, a quantum fluid in which superfluidity and
magnetism, both associated with symmetry breaking, are
simultaneously realized. $^{87}$Rb spinor condensates were rapidly
quenched across this transition to a ferromagnetic state and probed
using in-situ magnetization imaging to observe spontaneous symmetry
breaking through the formation of spin textures, ferromagnetic
domains and domain walls. The observation of topological defects
produced by this symmetry breaking, identified as polar-core
spin-vortices containing non-zero spin current but no net mass
current, represents the first phase-sensitive in-situ detection of
vortices in a gaseous superfluid.}

 Most ultracold atomic gases
consist of atoms with non-zero total angular momentum denoted by the
quantum number $F$, which is the sum of the total electronic angular
momentum and nuclear spin.  In spinor atomic gases, such as $F=1$
and $F=2$ gases of $^{23}$Na \cite{sten98spin,gorl03} and $^{87}$Rb
\cite{chan04,schm04},  all magnetic sublevels representing all
orientations of the atomic spin may be realized \cite{twocompref}.
The phase coherent portion of a Bose-Einsein condensed spinor gas is
described by a vector order parameter and therefore exhibits
spontaneous magnetic ordering. Nevertheless, considerable freedom
remains for the type of ordering that can occur. For $^{87}$Rb $F=1$
spinor gases, the spin-dependent energy per particle in the
condensate is the sum of two terms, $c_2 n \langle \vec{\hat{F}}
\rangle^2 + q \langle \hat{F}_z^2\rangle$, where $\vec{\hat{F}}$
denotes  the dimensionless spin vector operator. The first term
describes spin-dependent interatomic interactions, with $n$ being
the number density and $c_2 = (4 \pi \hbar^2/ 3 m) (a_2 - a_0)$
depending on the atomic mass $m$ and the $s$-wave scattering lengths
$a_{f}$ for collisions between pairs of particles with total spin
$f$ \cite{ho98,ohmi98}.  Given $c_2<0$ for our system
\cite{klau01rbspin,chan04,schm04,chan05coherent}, the interaction
term alone favors a ferromagnetic phase with broken rotational
symmetry. The second term describes a quadratic Zeeman shift in our
experiment, with $q = (h \times 70 \, \mbox{Hz}/\mbox{G}^2) B^2$ at
a magnetic field of magnitude $B$ \cite{zeemanfootnote}. This term
favors instead a scalar phase with no net magnetization, i.e.\ a
condensate in the $| m_z = 0\rangle$ magnetic sublevel. These phases
are divided by a second-order quantum phase transition at $q =
2|c_2| n$.

This article describes our observation of spontaneous symmetry
breaking in a $^{87}$Rb spinor BEC that is rapidly quenched across
this quantum phase transition.  Nearly-pure spinor Bose-Einstein
condensates were prepared in the scalar $|m_z = 0\rangle$ phase  at
a high quadratic Zeeman shift ($q\gg 2|c_2| n$). By rapidly reducing
the magnitude of the applied magnetic field, we quenched the system
to conditions in which the ferromagnetic phase is energetically
favored ($q\ll 2|c_2|n$). At variable times $\thold$ after the
quench, high-resolution maps of the magnetization vector density
were obtained using magnetization-sensitive phase contrast imaging
\cite{higb05larmor}. Soon after the quench, transverse ferromagnetic
domains of variable size formed spontaneously throughout the
condensate, divided by narrow unmagnetized domain walls separating
domains of nearly opposite orientation. Concurrent with the
formation of these domains, we also observed topological defects
that we characterize as single charged spin vortices with non-zero
circulation of spin current and an unmagnetized filled core.

Spinor BECs in the $|F=1, m_z = 0\rangle$ hyperfine state were
confined in an optical dipole trap characterized by oscillation
frequencies $(\omega_x, \omega_y, \omega_z) = 2 \pi (56, 350, 4.3)$
s$^{-1}$. The condensates, typically containing $2.1 (1) \times
10^6$ atoms, were formed at a magnetic field of 2 G and were
characterized by a peak density $n_0 = 2.8 \times 10^{14}$ cm$^{-3}$
and Thomas-Fermi radii $(r_x, r_y, r_z) = (12.8, 2.0, 167)$ $\mu$m
(Supplementary Note 1). The anisotropic condensates were effectively
two-dimensional with respect to the spatial variations in the
internal-state wavefunction due to the fact that the spin healing
length, $\xi_s = \sqrt{\hbar^2/2 m |c_2| n_0} = 2.4$ $\mu$m, is
larger than $r_y$, therfore eliminating spin dynamics in the
$\hat{y}$ direction. Thus, imaging the condensate in the $\hat{x} -
\hat{z}$ plane produces complete maps of the magnetization density.
The reduced dimensionality also determines the nature of topological
defects that may arise in this quantum fluid.

\begin{figure}
\centering

  \includegraphics[width=0.45\textwidth]{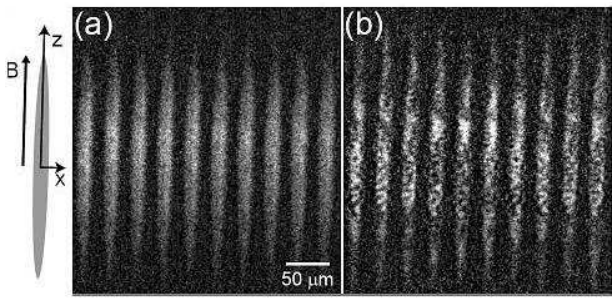}

  \caption{Direct imaging of inhomogeneous spontaneous magnetization of a spinor BEC.
  Transverse imaging sequences (first 10 of 24 frames taken) are shown (a) for
  a single condensate probed at $\thold = 36$ ms and
  (b) for a different condensate at $\thold = 216$ ms.  Shortly after the quench, the
  system remains in the unmagnetized $|m_z = 0\rangle$ state, showing neither short-range spatial
  nor temporal variation (i.e.\ between frames).  In contrast, condensates
  at longer times are spatially inhomogeneous and display
  spontaneous Larmor precession as
  indicated by the cyclical variation of signal strength vs.\ frame number.  Orientations of axes and of the
  magnetic field are shown at left.}
  \label{fig:imageset}

\end{figure}

After the condensate was formed, the magnetic field was ramped
linearly over 5 ms to a magnitude of 50 mG in the $\hat{z}$
direction and was held at this setting for a variable time $\thold$
prior to the imaging sequence. At this field, the quadratic Zeeman
energy $q = h \times 0.2 \, \mbox{Hz}$ is negligible compared to the
spin-dependent interaction energy of $(4/5) 2|c_2| n_0 = h \times
8.2 \, \mbox{Hz}$, which is computed using the density averaged in
the $\hat{y}$ direction.

The condensate magnetization was measured in-situ using
phase-contrast imaging with pulses of circular polarized light,
yielding an optical signal given approximately as $ \label{eq:lpsig}
\zeta \left(1+\frac{5}{ 6} \langle \hat{F}_y \rangle
+\frac{1}{6}\langle \hat{F}_y^2 \rangle\right) $
\cite{higb05larmor}. Here, $\zeta = \tilde{n} \sigma \gamma / 8
\delta$ where $\tilde{n}$ is the column density of the gas, $\sigma=
3 \lambda^2/2\pi$ is the resonant cross section, $\lambda$=795 nm is
the wavelength of the probe light, $\delta$ is the detuning from
resonance, and $\gamma$ is the natural linewidth (Supplementary Note
2). Thus, given the column density of the gas and ignoring the small
$\langle F_y^2\rangle$ signal, the phase-contrast signal is an
instantaneous measure of one component, $F_y = \langle \hat{F}_y
\rangle$, of the (dimensionless) magnetization of the gas.

We determined all three components of the vector magnetization
density with repeated images of the same atomic sample. First, a
sequence of 24 images was taken for which the magnetic field
remained oriented along the $\hat{z}$ direction (Supplementary Note
2). Larmor precession of the transverse magnetization at around 35
kHz is imaged as an aliased low-frequency oscillation (Fig. 1). The
complex transverse magnetization $F_T = F_x + i F_y$ can then be
determined from the relation $A(\vec{\rho}) \exp(i \phi(\vec{\rho}))
= i (5/6)\, \zeta(\rhovec)\, F_T(\rhovec)$ where $A(\vec{\rho})$ and
$\phi(\vec{\rho})$ are the amplitude and phase of this oscillation
at each pixel position $\vec{\rho}$. The normalization constant
$\zeta(\rhovec)$ was obtained by averaging the signal over all image
frames and fitting with a Thomas-Fermi distribution for the density
of the condensate. Following this sequence, two additional image
frames were obtained within 5 ms in which the magnetic field was
adiabatically reoriented in the $\hat{y}$ and $-\hat{y}$ directions.
The longitudinal magnetization was determined from the difference
between these last frames.

As shown in Fig. 1, at short times after the quench ($\thold < 50$
ms), the transverse magnetization images show no significant
variation across the cloud or between frames (similarly for the
longitudinal magnetization images, which are not shown). This
indicates that during this latency stage, the BEC remained in the
unmagnetized scalar phase (the  $|m_z=0\rangle$ state). Any
magnetization during this stage  either was too low in magnitude or
varied over too short a length scale to be discerned by our imaging.
At later times, a non-zero transverse magnetization signal
spontaneously developed, yielding a Larmor precession signal that
varied both in amplitude and in phase across the condensate. This
observation indicates the spontaneous, spatially inhomogeneous
breaking of $O(2)$ symmetry in the transverse plane in a direction
given by the phase of Larmor precession.

\begin{figure}
\centering

   \includegraphics[width=0.45\textwidth]{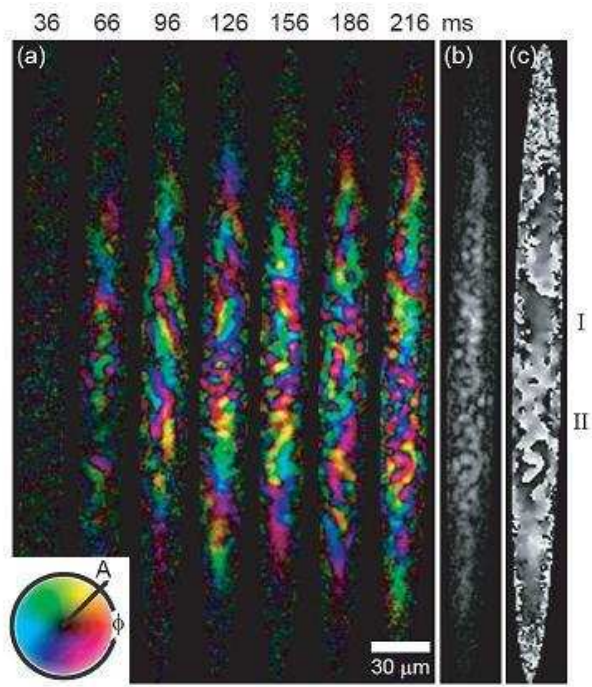}
  \caption{In-situ images of ferromagnetic domains and domain walls.  (a) The transverse
  magnetization density, measured for condensates at variable times $\thold$ (indicated on top),
   is shown with the magnetization
  orientation ($\phi = \arg({F}_T)$) indicated by hue, and magnitude ($A = \zeta |F_T|$)
  indicated by brightness.  The maximum brightness, indicated by the color wheel at left,
  corresponds to full magnetization of the condensate center.
  For the data at $\thold = 216$ ms, the magnetization density (b) and orientation (c) are
  shown separately, and regions with either a large (I)  or several small (II) magnetized domains are
  indicated.
The gray scale in (c) covers the range 0 to $2 \pi$.
   Regions outside the condensate are indicated in black.  Higher resolution picture available at
   http://physics.berkeley.edu/research/ultracold/pubs.html}
  \label{fig:phaseimages}
\end{figure}

In Fig. 2, the derived transverse magnetization for samples at
variable times $\thold$ are presented, rendered as color images to
portray both the magnetization orientation (as hue) and amplitude
(as brightness). These spatial maps show ferromagnetic domains of
variable size and orientation arising spontaneously following the
quench. The landscape of domains includes small regions of
homogeneous magnetization, with bands of unmagnetized gas, i.e.\
domain walls, separating regions of nearly opposite orientation. One
also observes larger ferromagnetic spin textures free of domain
walls in which the magnetization orientation varies smoothly over
tens of microns.

\textbf{Ferromagnetic domains: } The evolution toward ferromagnetism
from the $|m_z=0\rangle$ state under the conditions $q < |c_2|n$ is
a form of dynamical instability in a spinor BEC
\cite{pu99dyn,robi01instab,sait05spont,zhan05dyn}.  This instability
is a consequence of coherent collisional spin mixing in which pairs
of atoms in the $|m_z = 0 \rangle$ state collide inelastically to
produce atoms in the $|m_z = \pm 1\rangle$ states
\cite{chan04,schm04,wide05spin,chan05coherent}. Indeed, in our work,
measurements of populations in the different magnetic sublevels
through Stern-Gerlach analysis showed significant mixing of
populations coincident with the onset of spontaneous Larmor
precession.

One may also regard this instability as the phase separation of a
two-component condensate \cite{hall98dyn,mies99meta}, recalling that
the $|m_z = 0\rangle$ state represents an equal superposition of the
$|m_\phi = \pm 1\rangle$ eigenstates of any transverse spin
operator, $\hat{F}_\phi = \hat{F}_x \cos \phi + \hat{F}_y \sin\phi$.
Additionally, since $c_2 < 0$, the $\pm 1$ eigenstates of any spin
component are immiscible \cite{stam00leshouches}. Ferromagnetism
thus arises by the spinodal decomposition of a binary gaseous
mixture into neighboring regions of oppositely oriented transverse
magnetization. Applying results from the two-component case, this
phase separation is dominated by an instability with a
characteristic exponential timescale $\tinstab = \hbar / \sqrt{2
|c_2| n}$ with $n$ being the total gas density. The wavevector of
the dominant instability $\kinstab = \sqrt{2 m |c_2| n}/\hbar$,
defines the typical size $l = \pi \kinstab^{-1}$ of single-component
domains in the phase-separated fluid, and also the width $b \simeq
\kinstab^{-1}$ of domain walls in which the two components still
overlap \cite{gold97,timm98phas}.

To relate these predictions to our data, we considered the
density-weighted transverse magnetization correlation function
\begin{equation}
G_T(\delta \rhovec) = \mbox{Re} \left[ \frac{\sum_{\vec{\rho}}
\left(\zeta(\rhovec) F_T(\rhovec)\right)^* \left(\zeta(\rhovec +
\delta \rhovec) F_T(\rhovec + \delta \rhovec) \right)
}{\sum_{\vec{\rho}} \zeta(\rhovec) \zeta(\rhovec + \delta \rhovec)}
\right] \label{eq:defineg} \nonumber
\end{equation}
At zero range, $G_T(0)$ measures the degree to which the condensate
has evolved toward the ferromagnetic state.  As shown in Fig. 3 (a),
$G_T(0)$ rises after the quench from a near-zero background level
characteristic of the unmagnetized initial state of the condensate
to a nearly constant value of $G_T(0) \simeq 0.5$ during the
ferromagnetic stage.  This saturation value measures the area
occupied by domain walls.  Fitting $G_T(0)$ at early times ($\thold
< 90 \, \mbox{ms}$) to a rising exponential yields a time constant
of $\tau = 15(4) \, \mbox{ms}$. This value is in agreement with the
predicted $\tinstab = 13.7(3) \, \mbox{ms}$, calculated using the
peak density averaged over the $\hat{y}$ direction to account for
its reduced dimensionality. Over the same period of evolution, no
significant longitudinal magnetization was observed, confirming the
presence of purely transverse ferromagnetic domains.

\begin{figure}
\centering

  \includegraphics[width=0.45\textwidth]{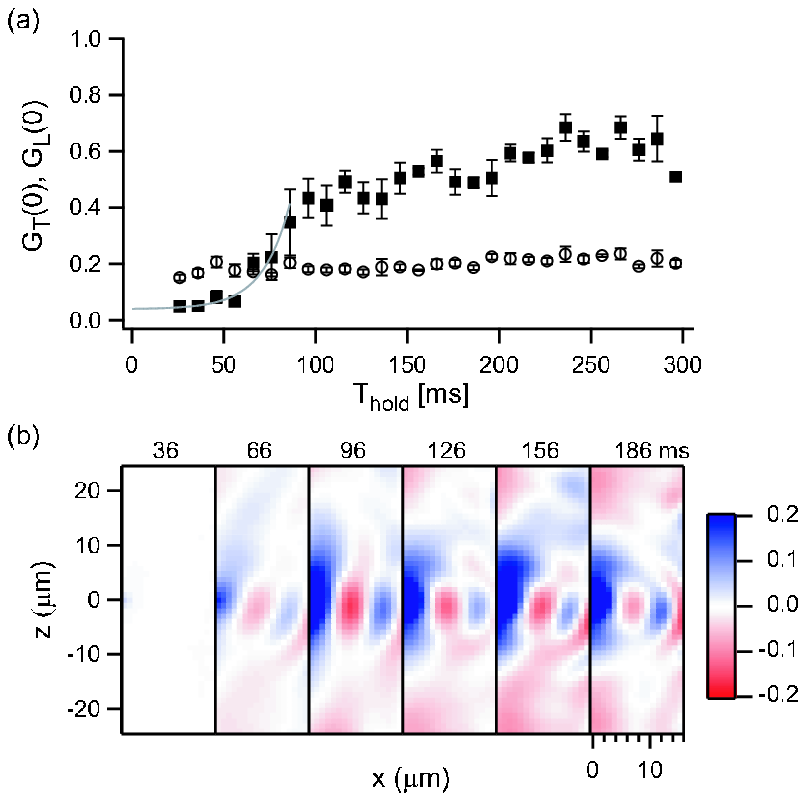}
  \caption{Temporal and spatial evolution of ferromagnetism in a quenched
  spinor BEC.   (a) The mean local squared transverse ($G_T(0)$, squares)
  and longitudinal ($G_L(0)$, circles) magnetization, averaged over several experimental
  repetitions (error bars indicate shot-to-shot rms fluctuations), is shown.  $G_T(0)$ rises
  exponentially with time constant $\tau = 15(4)\,
  \mbox{ms}$ as determined by a fit to data for $\thold <
  90\, \mbox{ms}$ (gray line), and saturates at $G_T(0)
  \simeq 0.5$ at later times.  No significant longitudinal magnetization
  is observed.  The minimum values
  for $G_T(0)$ and $G_L(0)$, evident in the data for earliest $\thold$, reflect residual
  noise due to processing either 24 or 2 frames, respectively, to obtain the magnetization density.
  (b) Maps of spatial correlations in the transverse magnetization ($G_T(x,z)$),
   at variable $\thold$ show alternating regions of positive and negative correlations in the narrow $\hat{x}$
   direction,
  indicating that
  phase separation occurs primarily along $\hat{x}$ via a few discrete dynamical instability modes.
At each $\thold,$ $G_T(x,z)$ was determined for ten repetitions of
the experiment and then averaged.
  }
  \label{fig:combined_correlations}
\end{figure}

Spatial correlations in the transverse magnetization (Fig. 3(b)) are
typified by a central region of positive correlations (near $\delta
\vec{\rho} = 0$) and then several equally-sized regions of
alternating negative and positive correlations displaced from one
another in the narrow ($\hat{x}$) dimension of the condensate.  We
estimate a typical size for single-component domains as  $\simeq 10
\, \mu$m, twice the displacement at which the transverse spin-spin
correlations changes sign, in good agreement with the predicted $\pi
\kinstab^{-1} = 8.3(2) \, \mu$m.  The presence of negative
correlation regions supports the model of spin-conserving
phase-separation discussed above.  The preferential phase separation
in the narrow $\hat{x}$ direction rather than along $\hat{z}$ is due
to the broader momentum distribution of the unmagnetized condensate
in that direction due to the finite condensate size. Thus, upon
quenching the system, a greater population of atoms is available to
seed the faster-growing, shorter-wavelength instabilities with
wavevector in the $\hat{x}$ direction.  The presence of several
alternations of positive and negative correlations  further suggests
that the phase separation occurs through a small number of discrete,
unstable magnetization modes.

\textbf{Topological defects: } The spontaneous symmetry breaking
observed in this work is one of many examples of symmetry breaking
that occur in nature. Symmetry breaking is presumed to have occurred
at thermal phase transitions in the early universe, giving rise to
the specific elementary particles and interactions observed in the
present day. An important aspect of rapid spontaneous symmetry
breaking, whether in the laboratory or of a cosmological nature, is
the creation of topological defects, a process described
theoretically by Kibble \cite{kibb76} and Zurek \cite{zure85cosmo}.
The types of topological defects that may be formed depend on the
group structure of the ground-state manifold reached at the
transition.

Analogues of cosmological topological defect formation have been
studied using liquid crystals \cite{chua91cosmo} and superfluid
helium \cite{hend94cosmo,ruut96,baue96}. In comparison with these
previous experiments, the present investigation focuses on a simpler
physical system, in which the time for thermal equilibration and
symmetry breaking is much longer than that needed to bring the
system across the symmetry-breaking transition. Furthermore, the
present work focuses on a quantum rather than a thermal phase
transition.

In our two-dimensional system, spin-vortices are topological point
defects  about which the orientation of magnetization has a $2 \pi
n$ winding with $n$ being a non-zero integer. Following a procedure
for identifying such defects in our data (Supplementary Note 3),
spin-vortex defects were observed with high confidence in about one
third of all images containing significant ferromagnetism ($\thold >
90$ ms), with some images containing up to four vortices.  Data
revealing one such spin-vortex are shown in Fig. 4. For this vortex,
the central region of near-zero Larmor precession amplitude has a
diameter of about 3 $\mu$m, comparable to the spin healing length
$\xi_s = 2.4 \, \mu$m defined earlier.  All the vortices we observed
were singly quantized, with no apparent preferred direction of
circulation.

\begin{figure}
\centering

  \includegraphics[width=0.35\textwidth]{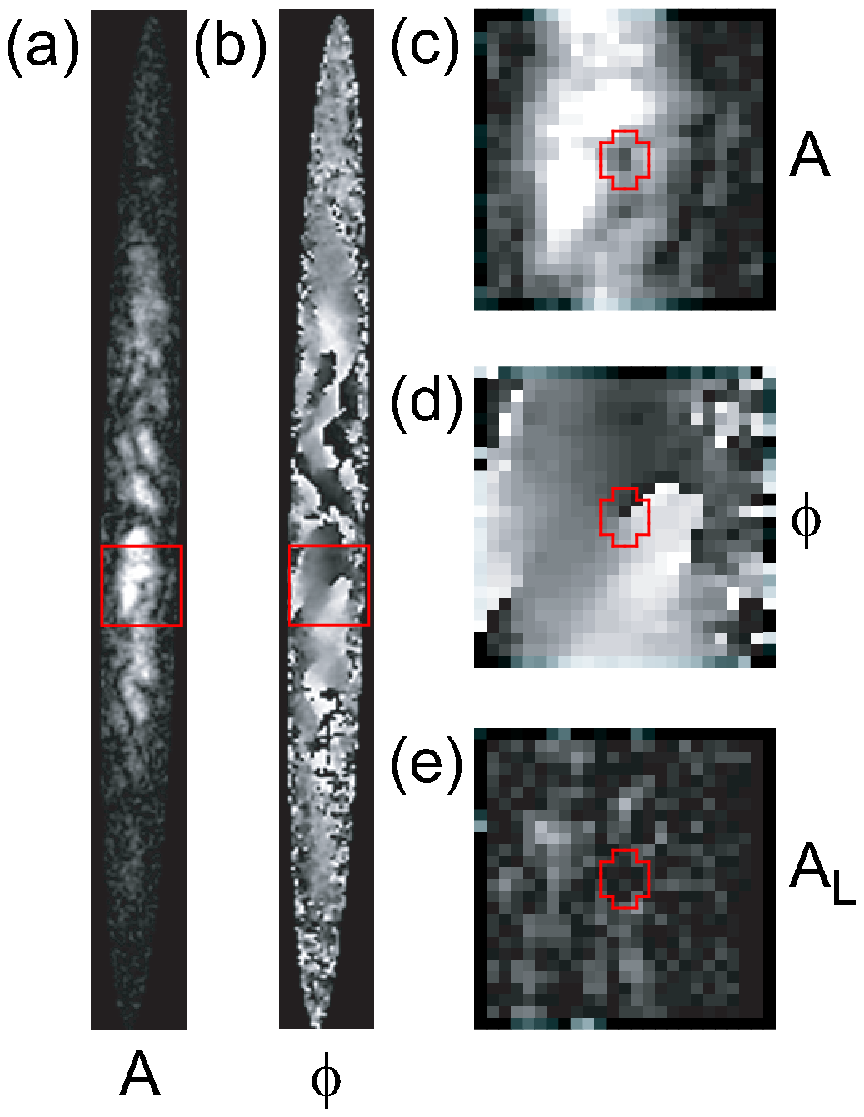}
  \caption{In-situ detection of a polar core spin-vortex.  Spatial maps
    of the transverse magnetization (a) magnitude ($A = \zeta
    |{F}_T|)$ and (b) orientation ($\phi = \arg({F}_T)$)
    are shown for a sample imaged at $\thold = 150$ ms.  Data from
    a portion
    of the condensate, indicated by boxes, are magnified, showing
    the transverse magnetization (c) magnitude and (d) orientation and also
   (e) the magnitude of the longitudinal magnetization ($A_L =
   \zeta |F_z|)$.  The phase along a closed path (indicated in red), surrounding
   a region of near-zero transverse magnetization and satisfying the criteria
   stated in the text, shows a net winding of $2 \pi$,
   revealing the presence of a spin-vortex defect with its core within the closed path.
   The core shows no significant longitudinal magnetization, allowing the identification
   of the defect as a polar core spin-vortex.  The gray scale for images (a,c,e) is the same.  The field of view is 31 $\times$ 320 $\mu$m for images (a,b),
   and 26 $\times$ 26 $\mu$m for images (c-e).
  }
  \label{fig:samplevortex}

\end{figure}

Many types of vortices that can occur in a gas with a
multi-component order parameter can be distinguished by the
composition of their cores \cite{isos01qvort}.  Based on our
measurements of the transverse and longitudinal magnetization at the
vortex core, the spin-vortices seen in our experiment appear to have
unmagnetized filled cores.  This observation supports their
characterization as ``polar-core'' spin-vortices (denoted as $(1, 0,
-1)$ vortices in Ref.\ \cite{isos01qvort}), for which the superfluid
order parameter is a superposition of atoms in the $|m_z = 1\rangle$
state rotating with one quantum of circulation, atoms in the $|m_z =
-1\rangle$ state rotating with one quantum of circulation in the
opposite sense, and non-rotating atoms in the unmagnetized $|m_z =
0\rangle$ state, which also fill the vortex core.  Such a vortex is
thus characterized by zero net mass circulation and a spin current
with one quantum of circulation. The origin for such spin currents
is presumably the spinodal decomposition by which ferromagnetism
emerges from the unmagnetized cloud.  The alternate identification
of these vortices as merons \cite{merm76, sait05chiral} is ruled out
by the absence of a longitudinal magnetization signal at the vortex
core.

In contrast with their topological nature in some other magnetic
systems, domain walls in a $F=1$ ferromagnet are not topologically
stable; rather, they may decay by the formation of spin
vortex-antivortex pairs.  As discussed above, we observe domain
walls that form at the onset of visible ferromagnetism in the gas,
permeate the inhomogeneous clouds, and persist for all times
thereafter. The persistence of domain walls in our system is a
subject for future investigation.

Of keen interest for future studies is the distinction between
thermal and quantum fluctuations as seeds for the unstable
symmetry-breaking process. In this work, the forced depletion of
atoms not in the $|m_z = 0\rangle$ state suggests that
ferromagnetism formed purely by the amplification of quantum
fluctuations, i.e.\ shot noise, a suggestion that warrants
experimental justification. Time-resolved experiments in which one
varies the rate at which the system is swept into the ferromagnetic
state may also uncover universal temporal dynamics that typify this
and other quantum phase transitions.  Our method of nondestructive
magnetization imaging also allows the behavior of a single quenched
condensate to be studied in detail, permitting an examination of the
dynamical evolution of domain walls and spin-vortices.

 We thank E.\ Mueller, J.\ Moore, and A.\
Vishwanath for helpful comments, J.\ Guzman for experimental
assistance, and the NSF and David and Lucile Packard Foundation
for financial support.  S.R.L.\ acknowledges support from the
NSERC.

\textbf{Supplementary Note 1: Experimental sequence}
Optically-trapped BECs used for this work were obtained by loading
around $10^8$ atoms in the $|F=1, m_z=-1\rangle$ state with
temperature of 2.5 $\mu$K into an optical dipole trap (ODT). The ODT
was formed by a single focus laser beam with a wavelength of 825 nm
that is linearly polarized to ensure that all components experience
the same trap potential. The weak quadratic dependance of the trap
depth on $m_F$ is negligible.  After loading the atoms, all were
placed in the $|m_z=0\rangle$ magnetic sublevel using rf rapid
adiabatic passage followed by application of a transient magnetic
field gradient of 4 G/cm.  Holding the magnetic field at a magnitude
of 2 G, the optical trapping beam was decreased in power over 400
ms, lowering the trap depth to $k_B \times$ 350 nK and the
temperature of the gas to $\sim$ 40 nK, well below the Bose-Einstein
condensation temperature. This yielded nearly-pure condensates of
$2.1(1) \times 10^6$ atoms.

After forming the condensate, the magnetic field was ramped within
5 ms to a value of 50 mG oriented in the $\hat{z}$ direction. The
field was held at this setting for a variable time before the
imaging sequence. Field gradients along the $\hat{x}$ and
$\hat{z}$ directions were nulled to less than 0.2 mG/cm. A
Stern-Gerlach analysis of populations in each of the magnetic
sublevels was applied to establish that the field ramp was
sufficiently slow as not to alter the spin state of the
condensate. Such an analysis bounded the populations in each of
the $|m_z = \pm 1 \rangle$ spin states to be less than 0.3\% of
the total population, both before and right after the field ramp.

\textbf{Supplementary Note 2:  Magnetization imaging} The probe
light for the phase contrast images was derived from a diode laser
detuned $\delta$ = -200 MHz from the $5S_{1/2} (F=1) \rightarrow
5P_{1/2} (F'=2)$ transition in $^{87}$Rb. The imaging sequence to
determine the transverse magnetization consisted of a sequence of
24 frames for which the probe was pulsed at a strobe frequency of
around 10 kHz. The duration of each individual pulse was around 1
$\mu$s.

\textbf{Supplementary Note 3: Image Analysis} The determination of
spin vortices in the magnetization images relied on the
identification of a core of "dark" pixels, consistent with zero
Larmor precession amplitude, which were surrounded by "bright"
pixels with a finite amplitude and well-defined phase of Larmor
precession. We identify vortices based on two criteria: (1) that
there is an island of dark pixels, a candidate for the vortex core,
which is surrounded entirely by bright pixels and that is at least
two bright pixels away from nearby dark pixels, and (2) that the
transverse magnetization traced along a closed loop through bright
pixels surrounding the core have a non-zero net winding. The
distinction between bright and dark pixels (at about one quarter of
the maximum Larmor precession amplitude) was chosen so as to
eliminate false-positive vortex detections in simulated data taking
into account the measured noise and resolution of our imaging
system, and using estimates for the width of contiguous domain
walls.


\begin{thebibliography}{10}
\bibitem{zeemanfootnote}
The linear Zeeman effect may be neglected due to conservation of
magnetization
  in our system.
\end{thebibliography}
\end{document}